\documentclass{vgtc}                          % final (conference style)
\graphicspath{{figures/}{pictures/}{images/}{./}} % where to search for the images

\usepackage{times}                     % we use Times as the main font
\usepackage{tabu}                      % only used for the table example
\usepackage{booktabs}                  % only used for the table example
\usepackage{lipsum}                    % used to generate placeholder text
\usepackage{mwe}                       % used to generate placeholder figures

\usepackage{mathptmx}                  % use matching math font
\usepackage{amsmath}
\usepackage{amssymb}

\onlineid{0}

\vgtccategory{Research}

\vgtcinsertpkg

\title{Scalable Semantic Steering of Embedding Projections}

\author{
Wei Liu\thanks{e-mail: wliu3@vt.edu, ORCID: \href{https://orcid.org/0009-0009-6340-8912}{0009-0009-6340-8912}}\\
\scriptsize Virginia Tech
\and
Eric Krokos\thanks{e-mail: ericpkrokos@gmail.com, ORCID: \href{https://orcid.org/0000-0003-1350-5297}{0000-0003-1350-5297}}\\
\scriptsize Department of Defense
\and
Kirsten Whitley\thanks{e-mail: visual.tycho@gmail.com, ORCID: \href{https://orcid.org/0000-0003-1356-326X}{0000-0003-1356-326X}}\\
\scriptsize Department of Defense
\and
Rebecca Faust\thanks{e-mail: rfaust1@tulane.edu, ORCID: \href{https://orcid.org/0000-0002-7640-1287}{0000-0002-7640-1287}}\\
\scriptsize Tulane University
\and
Chris North\thanks{e-mail: north@vt.edu, ORCID: \href{https://orcid.org/0000-0002-8786-7103}{0000-0002-8786-7103}}\\
\scriptsize Virginia Tech
}

\teaser{
  \centering
  \includegraphics[width=\linewidth]{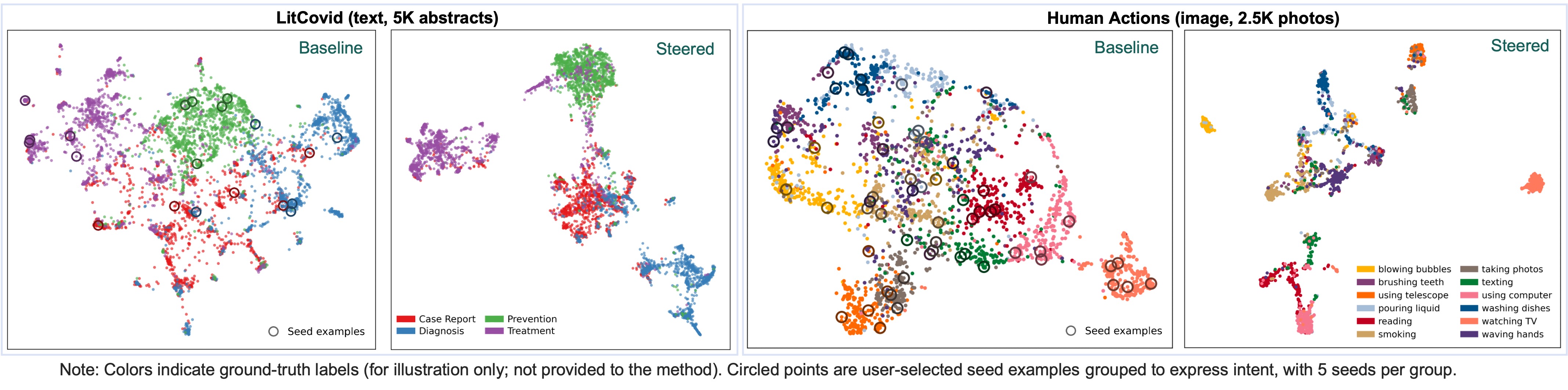}
  \caption{Prototype-based semantic steering on a text (LitCovid) dataset and an image (Human Actions) dataset. In each pair, the baseline projection (left) is reorganized into a more semantically coherent layout (right) using five user-selected seed examples per group (circled). Our method propagates analyst intent across the full collection through embedding-space operations driven by a single LLM call.}
  \label{fig:teaser}
}

\abstract{
Low-dimensional projections support interactive visual analysis of high-dimensional data embeddings, but their structure often does not align with analyst-defined semantic relationships. Recent LLM-augmented semantic steering methods address this gap by externalizing analyst intent from user-defined groups of seed examples, but they propagate intent through per-item LLM reasoning, causing LLM calls and cost to grow linearly with collection size. We propose a scalable semantic steering method that shifts semantic computation from individual items to user-defined groups. A single LLM call generates structured profiles for all groups, which are embedded and combined with seed centroids to form hybrid semantic prototypes. The method then propagates intent without retraining, using embedding-space soft assignment, abstention, and alignment-scaled updates before reprojection. On a 5K-document LitCovid corpus, our method achieves global alignment comparable to per-item LLM steering while reducing LLM calls by over three orders of magnitude. An image case study shows that the same prototype-based mechanism extends to multimodal embeddings. These results suggest that group-level representations can make semantic steering more practical for larger embedding collections.
} % end of abstract


\begin{document}


\firstsection{Introduction}

\maketitle

%% \section{Introduction} %for journal use above \firstsection{..} instead
Low-dimensional projections of high-dimensional embeddings are widely used to support visual analysis of text and image collections, enabling analysts to explore semantic structure through spatial organization \cite{huang2023va+, liu2024visualizing}. However, projection structure is largely determined by the underlying embedding model and dimensionality reduction technique, which may not align with the semantic relationships that analysts intend to examine for a given task \cite{lin2024imagesi, bian2021deepsi}. Semantic interaction (SI) \cite{endert2012semantic} addresses this gap by allowing analysts to reshape projections through spatial actions such as grouping or repositioning items. More recent work augments SI with large language models (LLMs) to externalize analyst intent as natural-language semantics rather than as implicit model parameters, propagating it across the dataset through item-level reasoning \cite{liu2026llmaugmented, oliveira2025creating}. 

While effective, this design has a fundamental scalability bottleneck: semantic propagation is performed \emph{per item}. Given a small set of seed examples, an LLM is invoked for every remaining item to evaluate its relationship to the analyst's intent and generate item-level augmentations. The cost grows linearly with collection size. On a 5{,}000-item corpus, a single steering interaction issues over 5{,}000 LLM calls and costs roughly \$32 under current pricing~\cite{openai_api} (Section~\ref{sec:rq2}). As datasets grow, this design becomes increasingly impractical for interactive use.

We make a structural observation: in SI, analysts often express intent at the group level, not at the item level. Users specify a few example groupings, and the relationships of interest live between these groups. This observation is reinforced by two parallel developments outside the visual analytics community. In few-shot learning, a class can be represented by a prototype computed as the centroid of a small support set~\cite{snell2017prototypical}. Recent vision–language work shows that textual class descriptions can serve as semantic prototypes~\cite{radford2021learning, pratt2023does}, and that combining text-derived prototypes with support-set centroids further strengthens them~\cite{palanisamy2024proto, goswami2026cross}. Translated to SI, these insights suggest that semantic intent can be externalized \textbf{once per group as a hybrid prototype}: the seed centroid for \emph{data-grounded local geometry}, and the LLM-generated profile embedding for \emph{language-articulated abstraction} that generalizes beyond seeds. Intent can then be propagated entirely through embedding-space operations, removing the need for per-item LLM reasoning.

Building on this insight, we propose a scalable semantic steering method based on group-level semantic prototypes. From user-defined groups, a single LLM call generates structured semantic profiles for all groups jointly. Each profile is embedded and combined with its seed centroid into a hybrid prototype. Items are softly assigned to prototypes via embedding similarity and updated through alignment-scaled blending before reprojection.  The entire pipeline issues one LLM call regardless of dataset size; the method operates purely in embedding space and extends naturally to multimodal data through encoders such as CLIP \cite{radford2021learning} without retraining. This work makes the following contributions:

(1) \textbf{A scalable method} for LLM-augmented semantic steering that shifts semantic computation from per-item reasoning to group-level abstraction, reducing LLM cost from O(N) to O(1).

(2)  \textbf{A hybrid prototype design} fusing seed centroids with LLM-generated group-profile embeddings, propagated through adaptive soft assignment and alignment-scaled embedding updates.

(3)  \textbf{Evidence on a 5K text corpus} that the method achieves similar alignment quality to per-item LLM steering at over $1000\times$ lower cost, with a case study showing the method extends to images.

\section{Related Work}
\noindent\textbf{Semantic Interaction and Projection Steering.} SI lets analysts shape projection layouts through spatial interactions such as grouping or repositioning~\cite{endert2012semantic, sacha2016visual, endert2012semantics}. Such signals are typically translated into geometric constraints, feature reweighting, or learned distance metrics~\cite{brown2012dis, self2018observation, dowling2018sirius}, and more recent frameworks extend SI to deep neural encoders by fine-tuning embeddings~\cite{bian2021deepsi, lin2024imagesi}. While effective, intent is encoded implicitly through model parameters, requiring per-interaction optimization that scales poorly to larger collections.

\noindent\textbf{LLM-Based Steering with Explicit Semantics.} 
Recent work uses LLMs to externalize analyst intent as natural-language semantics. Oliveira et al.~\cite{oliveira2025creating} steer projections via zero-shot classification against user-specified categories, requiring the full label set in advance. Liu et al.~\cite{liu2026llmaugmented} remove this requirement by deriving ``cluster cards'' from grouped examples, but issue a per-item LLM call for selective augmentation. ModalChorus \cite{ye2024modalchorus} aligns multimodal embeddings via spatial drag operations and backend fine-tuning. A complementary line of work uses LLMs to describe existing clusters, leaving the layout unchanged \cite{buchmuller2026langlasso}. These steering methods require per-item LLM reasoning or iterative model fine-tuning, creating a fundamental scalability bottleneck for interactive use.

\noindent\textbf{Prototype-Based Representations.} Two lines of work inform our design. Prototypical Networks~\cite{snell2017prototypical} represent each class as the mean embedding of its support examples for few-shot classification. In vision–language and text models, class names or descriptions can themselves act as prototypes in a shared embedding space \cite{radford2021learning, hong2022empowering, pratt2023does}, and recent few-shot methods strengthen prototypes by incorporating textual information alongside support-set centroids \cite{palanisamy2024proto, pahde2021multimodal, goswami2026cross}. 
These advances target classification; we adapt the prototype formulation to semantic steering, fusing each user-defined group's seed examples and description into a hybrid prototype.

\section{Method}

We propose a scalable semantic steering method that replaces per-item LLM reasoning with group-level semantic prototypes. Given a small number of analyst-defined groups, our approach externalizes semantic intent in \textbf{a single LLM call} and propagates it across the entire collection through embedding-space operations. The pipeline is modality-agnostic and applies to both text and image collections.

\subsection{Overview and Problem Setup}
\label{sec:method-overview}

\begin{figure}[t]
    \centering
    \includegraphics[width=\columnwidth]{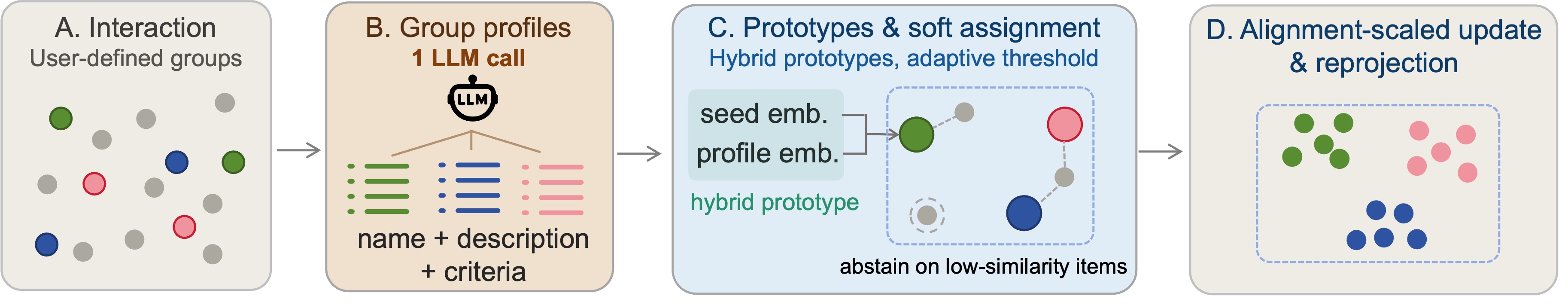}
    \caption{ Scalable semantic steering pipeline. From user-defined groups (A), a single LLM call generates structured profiles for all groups jointly (B). Profiles are embedded and combined with seed centroids to form hybrid prototypes; items are softly assigned to prototypes via embedding similarity, with low-similarity items left unassigned (C). Embeddings are updated with alignment-scaled blending and reprojected (D). Only stage B involves an LLM call, making the total LLM cost independent of dataset size.
    }
    \label{fig:interaction-propagation}
\end{figure}

Let $\mathcal{D} = \{d_i\}_{i=1}^{N}$ denote a collection of items (documents or images) with embeddings $\{z_i \in \mathbb{R}^d\}$ produced by a pretrained encoder. A baseline projection is obtained by applying a dimensionality reduction method (e.g., UMAP~\cite{mcinnes2018umap}) to $\{z_i\}$. During an interaction, the analyst selects a small number of items and organizes them into $K$ groups $\{G_k\}_{k=1}^{K}$, each expressing a semantic relationship of interest. Our goal is to reshape the projection to reflect the analyst's intent through group-level semantic abstraction, with LLM usage independent of $N$. All embeddings are $L_2$-normalized; $\operatorname{normalize}(\cdot)$ denotes renormalization after vector combinations. Figure~\ref{fig:interaction-propagation} illustrates the pipeline: we generate group profiles (\S\ref{sec:method-profile}), construct hybrid prototypes (\S\ref{sec:method-prototype}), propagate intent through soft assignment with abstention (\S\ref{sec:method-assignment}), and update embeddings before reprojection (\S\ref{sec:method-update}).

\subsection{Group Profile Generation}
\label{sec:method-profile}
We issue one LLM call that returns profiles for all $K$ groups---each profile contains a name, a short description, and inclusion and exclusion criteria. This adopts the cluster card formulation from prior LLM-augmented steering work~\cite{liu2026llmaugmented}, but it is the \emph{only} LLM call our method requires. Each profile is then encoded with the same encoder used for items, yielding a profile embedding $p_k^{\text{profile}} \in \mathbb{R}^d$ per group. For image collections, we use a multimodal encoder (e.g., CLIP~\cite{radford2021learning}) to embed text profiles into the shared image-text space, enabling cross-modal similarity in subsequent stages.

\subsection{Hybrid Semantic Prototypes}
\label{sec:method-prototype}

Each group is represented by a hybrid prototype that combines two complementary signals. The \emph{seed centroid} $p_k^{\text{seed}} = \operatorname{normalize}\!\left( \frac{1}{|G_k|} \sum_{i \in G_k} z_i \right)$ grounds the prototype in concrete examples provided by the analyst, preserving local geometry of the original embedding space. The \emph{profile embedding} $p_k^{\text{profile}}$ injects high-level semantic abstraction articulated in natural language, which generalizes beyond the specific seed instances. The hybrid prototype is their convex combination:
\begin{equation}
p_k = \operatorname{normalize}\!\left(\gamma\, p_k^{\text{seed}} + (1-\gamma)\, p_k^{\text{profile}}\right),
\end{equation}
where $\gamma \in [0, 1]$ controls the trade-off. Setting $\gamma = 1$ recovers seed-only steering; $\gamma = 0$ uses only the LLM-articulated semantics. We use $\gamma = 0.5$ by default and study its effect in Section~\ref{sec:experiments}.

\subsection{Soft Assignment with Adaptive Threshold}
\label{sec:method-assignment}

We propagate semantic intent to the full collection by computing similarities between items and prototypes, $s_{ik} = \cos(z_i, p_k)$. Instead of a global threshold, we compute a \emph{per-group adaptive threshold} that calibrates each prototype against the analyst's seeds. Let $D_k^+ = G_k$ be the seeds of group $k$ and $D_k^- = \bigcup_{j \neq k} G_j$ the seeds of all other groups. We define $b_k = \tfrac{1}{2}\!\left(\bar{s}_k^{+} + \bar{s}_k^{-}\right)$, where $\bar{s}_k^{+}$ and $\bar{s}_k^{-}$ are the mean similarities of in-group and out-of-group seeds to $p_k$, respectively. This threshold automatically locates a midpoint between what the analyst considers ``in'' and ``out'' for each group. Soft weights are then computed as
\begin{equation}
w_{ik} = \sigma\!\left(\tau (s_{ik} - b_k)\right),
\end{equation}
where $\sigma(\cdot)$ denotes the logistic sigmoid function and $\tau$ controls sharpness ($\tau = 20$ by default).

To preserve selectivity, we apply two filters. First, an \emph{adaptive top-$k$ mask} $\mathcal{K}_i = \left\{k \;:\; w_{ik} \geq \rho \cdot \max_{j} w_{ij}\right\}$ retains only groups whose weight is close to the maximum (with $\rho = 0.85$). This allows borderline items to remain associated with multiple prototypes while assigning most items primarily to a single group. Second, an \emph{abstain gate} leaves items unchanged when $\max_k w_{ik} < t_{\text{none}}$, mirroring the abstention mechanism in per-item LLM steering~\cite{liu2026llmaugmented} but using embedding similarity rather than language reasoning.

\subsection{Alignment-Scaled Update and Reprojection}
\label{sec:method-update}

For each non-abstained item, we compute a steering target as the weight-normalized combination of its assigned prototypes:

\begin{equation}
m_i = \operatorname{normalize}\!\left( \frac{\sum_{k \in \mathcal{K}_i} w_{ik}\, p_k}{\sum_{k \in \mathcal{K}_i} w_{ik}} \right).
\end{equation}

The updated embedding is an alignment-scaled interpolation between the original embedding and the steering target:
\begin{equation}
z_i' = \operatorname{normalize}\!\big( (1 - \lambda_i)\, z_i + \lambda_i\, m_i \big), \qquad \lambda_i = \alpha \cdot \max_{k} w_{ik},
\end{equation}

where $\alpha \in [0, 1]$ is a global steering strength and $\max_k w_{ik}$ scales the blend by per-item prototype alignment. Unlike prior work that applies a fixed blending coefficient uniformly across all steered items~\cite{liu2026llmaugmented}, our update strength is \textbf{per-item}: items with strong alignment to a prototype are pulled more strongly toward the corresponding target, while ambiguous items move only slightly. Abstained items use $\lambda_i = 0$. The updated embeddings $\{z_i'\}$ are reprojected with the same dimensionality reduction method as the baseline.

\section{Experiments}
\label{sec:experiments}

We evaluate whether group-level semantic prototypes can match the steering quality of per-item LLM reasoning at substantially lower cost.

\subsection{Setup}
\label{sec:exp-setup}

\hspace*{\parindent} \textbf{Dataset.} We use a 5{,}000-article subset of the LitCovid corpus~\cite{chen2022multi}, evenly distributed across four categories: Case Report, Diagnosis, Prevention, and Treatment (1{,}250 articles each).

\textbf{Embeddings and projection.} Text documents are embedded using \texttt{text-embedding-3-small}~\cite{openai_api} and projected to 2D with UMAP~\cite{mcinnes2018umap} (cosine distance, $n_{\text{neighbors}}{=}30$, $\text{min\_dist}{=}0.1$). 

\textbf{Simulated interaction.} To enable controlled and reproducible evaluation, we simulate analyst interaction by sampling 5 seed items per ground-truth category and treating them as user-defined groups. Reference labels are used only for evaluation and are never exposed to the LLM or the steering pipeline. All results are averaged over 3 random seed selections and reported as mean $\pm$ std.

\textbf{Metrics.} We report three groups of metrics relative to the unsteered baseline. \textbf{(1) Alignment:} $\Delta$Sil measures global separation using a scaled silhouette score (Sil $= 2s$, where $s$ is the standard silhouette~\cite{rousseeuw1987silhouettes}; Sil $\approx 1$ is treated as ideal, indicating well-separated but not overly compact groups~\cite{lin2024imagesi}). $\Delta$NC measures local coherence as the average fraction of same-group items among each point's $k=10$ nearest neighbors in the 2D projection \cite{liu2026llmaugmented}. \textbf{(2) Assignment behavior:} top-1 accuracy from $\arg\max_k w_{ik}$; top-2 set-hit accuracy (whether the true group appears among the two highest-weighted prototypes); abstain rate (fraction with $\max_k w_{ik} < t_{\mathrm{none}}$). Accuracy is reported as \emph{full} (counting abstained items as incorrect) and \emph{assigned} (computed over non-abstained items only). \textbf{(3) Cost:} total LLM calls and dollar cost, computed under GPT-5.4 pricing \cite{openai_api}.

\textbf{Compared methods.} We use the unsteered projection as a baseline and compare four steering configurations: \textbf{per-item LLM steering}~\cite{liu2026llmaugmented}, which performs LLM reasoning for each non-seed item, and three variants of our method---\textbf{profile-only} ($\gamma{=}0$), \textbf{hybrid} ($\gamma{=}0.5$), and \textbf{seed-only} ($\gamma{=}1$). 
Since $\alpha_{\text{ours}}$ upper-bounds the item-specific blend strength, whereas per-item LLM steering applies a single coefficient across items, we set the per-item LLM steering coefficient to our method’s mean effective blend strength over steered items for a fair comparison.

\textbf{Implementation.} All LLM calls use \texttt{gpt-5.4}~\cite{openai_api} with temperature $0$. By default, we use $\gamma{=}0.5$, $\tau{=}20$, $\rho{=}0.85$, and $t_{\text{none}}{=}0.20$, and $\alpha_{\text{ours}}=0.9$.

% ---------------- RQ1 ----------------
\subsection{RQ1: Does group-level prototype steering match the alignment quality of per-item LLM steering?}
\label{sec:rq1}
Table~\ref{tab:litcovid_alignment} reports alignment results on LitCovid. Our hybrid prototype achieves the strongest global alignment ($\Delta\mathrm{Sil} = 0.302$), comparable to per-item LLM steering ($0.259$) despite issuing no per-item LLM calls. This suggests that a small number of group-level prototypes carry enough semantic information to globally reorganize a 5{,}000-document collection. The seed-only variant achieves lower but still comparable global alignment ($0.248$), while profile-only is noticeably weaker ($0.190$), indicating that data-grounded structure and language-mediated abstraction contribute complementary signals; we examine this further in Section~\ref{sec:rq3}.

Per-item steering retains an advantage in local consistency ($\Delta\mathrm{NC} = 0.071$ vs.\ $0.023$). This gap reflects a structural difference between the two approaches: per-item reasoning produces a unique augmentation per document, allowing each to shift along a document-specific direction, whereas prototype-based steering pulls all documents in a group toward a shared target. The latter strengthens between-group separation but reduces within-group differentiation. A similar pattern appears in assignment behavior: per-item reaches higher top-1 accuracy on assigned items ($0.90$ vs.\ $0.79$), but our top-2 set-hit accuracy ($0.94$) approaches per-item's top-1, indicating that disagreements primarily concern ambiguity between adjacent groups rather than misclassification (abstain rates are comparable: $0.110$ vs.\ $0.115$). Overall, group-level prototypes match per-item LLM steering on global alignment, with a controlled trade-off on within-group local structure.

\begin{table}[t]
\centering
\caption{Alignment quality on LitCovid ($N=4{,}999^{\dagger}$, mean $\pm$ std over 3 random repetitions). 
$\Delta$Sil and $\Delta$NC are reported relative to the unsteered baseline (Sil=0.435, NC=0.779). 
Accuracy: full (assigned)$^{\ddagger}$.}
\label{tab:litcovid_alignment}
\small
\setlength{\tabcolsep}{3.8pt}
\renewcommand{\arraystretch}{0.92}
\begin{tabular}{lcccc}
\hline
\textbf{Method} & \textbf{$\Delta$Sil $\uparrow$} & \textbf{$\Delta$NC $\uparrow$} & \textbf{Acc top-1 $\uparrow$} & \textbf{Acc top-2 $\uparrow$} \\
\hline
Per-item LLM \cite{liu2026llmaugmented} $^\S$
& $0.259 \pm 0.015$ 
& $\mathbf{0.071} \pm 0.003$ 
& $\mathbf{0.80}$ ($\mathbf{0.90}$) 
& -- \\
Ours (seed-only) 
& $0.248 \pm 0.033$ 
& $0.010 \pm 0.012$ 
& $0.64$ ($0.78$) 
& $0.77$ ($\mathbf{0.94}$) \\
Ours (profile-only) 
& $0.190 \pm 0.035$ 
& $0.003 \pm 0.004$ 
& $0.71$ ($0.74$) 
& $\mathbf{0.87}$ ($0.91$) \\
\textbf{Ours (hybrid)} 
& $\mathbf{0.302} \pm 0.070$ 
& $0.023 \pm 0.006$ 
& $0.70$ ($0.79$) 
& $0.83$ ($\mathbf{0.94}$) \\
\hline
\end{tabular}

\vspace{0.5em}
\begin{minipage}{0.95\linewidth}
\footnotesize
$^{\dagger}$ One article excluded due to non-parseable LLM responses in the per-item baseline on some seeds. \\
$^{\ddagger}$ ``Full'' counts abstained items as incorrect; ``assigned'' computes over non-abstained items only. Abstain rates: 0.110 (per-item), 0.115 (ours, hybrid). Per-item steering produces hard assignments and has no top-2.
$^\S$ Per-item LLM uses blending coefficient $\bar{\lambda}_{\text{eff}} = 0.49 \pm 0.02$ (see \S\ref{sec:exp-setup}) 
\end{minipage}
\end{table}

% ---------------- RQ2 ----------------
\subsection{RQ2: How much computational cost is reduced?}
\label{sec:rq2}
A central motivation for our method is that semantic reasoning need not be performed for every item. By externalizing semantic intent once per group rather than per item, the cost of LLM reasoning becomes independent of dataset size.

Table~\ref{tab:computational-cost} quantifies this difference on LitCovid. At $N=5\mathrm{K}$, per-item LLM steering issues $5{,}001$ reasoning calls and costs \$32.35 per run, while our method issues a single LLM call (the group profile generation) at \$0.025 per run, a $1{,}294\times$ reduction in cost. Per-item cost scales linearly with $N$ while ours remains constant; the gap widens at larger scales. This shift has direct implications for interactive use. Per-item LLM steering becomes increasingly impractical as datasets grow, both in monetary cost and in the latency incurred by thousands of sequential or batched LLM calls. Our method, in contrast, requires only one LLM call regardless of $N$; subsequent stages reduce to embedding similarity and matrix operations. Together with the alignment results in Section~\ref{sec:rq1}, these findings indicate that group-level abstraction enables semantic steering to operate at an interactive scale.

\begin{table}[t]
\centering
\caption{Computational cost comparison. Per-item steering scales as $\mathcal{O}(N)$; ours issues one LLM call regardless of $N$. $N=5\text{K}$ values are measured as the mean across 3 runs; other rows are projected.}
\label{tab:computational-cost}
\small
\setlength{\tabcolsep}{4pt}
\begin{tabular}{@{}lrrrrrrr@{}}
\toprule
 & \multicolumn{3}{c}{Per-item LLM \cite{liu2026llmaugmented}} & \multicolumn{3}{c}{Ours} & \\
\cmidrule(lr){2-4} \cmidrule(lr){5-7}
$N$ & calls & tokens (in\,/\,out) & \$ & calls & tokens (in\,/\,out) & \$ & ratio \\
\midrule
500\textsuperscript{\dag}  & 501     & 0.80M\,/\,80K  & 3.21  & 1 & 6.4K\,/\,0.59K & 0.025 & 128$\times$ \\
1K\textsuperscript{\dag}   & 1{,}001 & 1.61M\,/\,161K & 6.45  & 1 & 6.4K\,/\,0.59K & 0.025 & 258$\times$ \\
2K\textsuperscript{\dag}   & 2{,}001 & 3.24M\,/\,322K & 12.92 & 1 & 6.4K\,/\,0.59K & 0.025 & 517$\times$ \\
5K                          & 5{,}001 & 8.10M\,/\,807K & 32.35 & 1 & 6.4K\,/\,0.59K & 0.025 & 1294$\times$ \\
\bottomrule
\end{tabular}

\vspace{2pt}
\footnotesize
\textsuperscript{\dag}\,Projected; per-item cost scales as $N \times$ per-call cost (measured at $N=5$K). GPT-5.4 pricing: \$2.50/M input, \$15/M output \cite{openai_api}. 
\end{table}

% ---------------- RQ3 ----------------
\subsection{RQ3: What roles do seed and profile signals play?}
\label{sec:rq3}

Our prototype design combines two complementary signals: the centroid of seed examples ($\gamma=1$) and the embedding of the LLM-generated group profile ($\gamma=0$). Table~\ref{tab:litcovid_alignment} reports three variants.

The \emph{seed-only} variant grounds prototypes directly in the analyst's chosen examples, preserving local structure of the original embedding space. It achieves moderate global alignment ($\Delta\mathrm{Sil} = 0.248$) but the lowest top-1 accuracy ($0.64$), suggesting that seed centroids alone do not generalize beyond the immediate neighborhood of the chosen examples. The \emph{profile-only} variant relies entirely on language-mediated abstraction: the LLM articulates the shared meaning of each group, and this articulation drives prototype placement. It achieves the lowest global alignment ($\Delta\mathrm{Sil} = 0.190$) but the lowest abstain rate ($0.051$) and competitive top-1 accuracy on assigned items ($0.74$), indicating that abstract semantic descriptions allow many documents to be assigned, but lack the data grounding needed to reorganize the projection at scale. The \emph{hybrid} variant ($\gamma=0.5$) combines both signals and achieves the strongest global alignment ($\Delta\mathrm{Sil} = 0.302$). The two signals are complementary: seed centroids provide data-grounded local geometry, while profile embeddings provide language-articulated abstraction that generalizes beyond the seed neighborhood. Neither alone reproduces the hybrid's performance, supporting our design choice of combining both signals instead of relying on either one alone.

% ---------------- RQ4 ----------------

\section{Case Studies}

We illustrate prototype-based steering on the two datasets in Figure ~\ref{fig:teaser}, focusing on how user-defined groups reshape the projection and on how the method generalizes to both text and images.

\textbf{Text: LitCovid.} The LitCovid baseline projection (Figure~\ref{fig:teaser}, left pair) places articles primarily by the geometry of their language-model embeddings. The four COVID-19 aspects overlap heavily in the central region: \emph{Prevention} partially separates at the top, but \emph{Diagnosis}, \emph{Treatment}, and \emph{Case Report} remain intermixed, making aspect structure difficult to read off the layout. Given five seed articles per aspect, our hybrid approach ($\gamma = 0.5$) reorganizes the layout into four spatially separated regions consistent with the analyst's intent. \emph{Treatment} articles consolidate into a single cohesive cluster on the left, and \emph{Diagnosis} and \emph{Case Report} (two aspects the baseline conflates) separate into distinct neighborhoods. The visual reorganization corresponds to the hybrid row of Table~\ref{tab:litcovid_alignment} and is qualitatively comparable to per-item LLM steering~\cite{liu2026llmaugmented}, obtained here without any per-item LLM inference.

\textbf{Images: Stanford-40 Actions.}
To examine whether group-level abstraction generalizes beyond text, we apply the same pipeline to image embeddings on a 12-category subset of Stanford-40 Actions~\cite{yao2011human} (2{,}583 images across 12 action categories). Image embeddings come from CLIP ViT-B/32~\cite{radford2021learning}, and group profiles are generated by GPT-5.4 \cite{openai_api} in vision mode using a JSON schema similar to that used in the text pipeline. UMAP uses \texttt{n\_neighbors} $= 15$; all other settings match the text pipeline.

The baseline CLIP projection (Figure~\ref{fig:teaser}, right pair) forms a single connected manifold in which most categories overlap in the central region. After steering, the layout reorganizes into a set of well-separated regions aligned with the user-defined groups. Several categories that were heavily entangled in the baseline, such as \emph{waving hands} and \emph{brushing teeth}, form distinct neighborhoods in the steered projection, while a smaller residual mixed region remains where visually similar actions (e.g., \emph{texting} and \emph{pouring liquid}) are not fully resolved. Quantitatively, the steered projection achieves $\Delta\mathrm{Sil} = 0.20$, $\Delta\mathrm{NC} = 0.05$, and top-1 accuracy $= 0.70$ and top-2 accuracy $= 0.83$ on non-seed images. These results suggest that the method extends naturally to other modalities with a compatible joint embedding space and a group-level profiling mechanism.

\section{Discussion}

\hspace*{\parindent}\textbf{From per-item reasoning to group-level abstraction.} Our results suggest that the scalability bottleneck in LLM-augmented
semantic steering is partly a mismatch in granularity. Prior methods~\cite{liu2026llmaugmented, oliveira2025creating} reason about each item because projections display individual points, but
analysts often express intent by selecting and grouping examples~\cite{bian2021deepsi, endert2012semantics}. By shifting semantic reasoning from items to user-defined groups, our method aligns the computational unit with the interaction unit. The LLM externalizes intent once as group profiles; the method then uses the resulting prototypes to propagate that intent through embedding-space operations.

\textbf{Overview steering versus local refinement.}
Prototype steering is best viewed as a scalable overview-level steering mechanism for exploratory sensemaking. Because items are moved toward shared group-level prototypes, the method strengthens between-group separation but provides less item-specific adaptation than per-item LLM steering~\cite{liu2026llmaugmented}. This makes the method suitable for reorganizing a projection around a small number of semantic concepts and obtaining a clearer overview of how a collection aligns with analyst intent. Tasks requiring fine-grained local neighborhoods or subtle within-group distinctions may benefit from subsequent local interactions or per-item refinement. 
The generated profiles also provide inspectable group-level semantics, although prototype assignments do not provide item-level rationales as in per-item LLM steering.

\textbf{Limitations and future work.}
Our quantitative evaluation uses simulated groups derived from ground-truth labels, but real analyst-selected groups may be noisier, more ambiguous, or less representative \cite{wenskovitch2024toward}. Performance also depends on profile quality: incorrect or overly broad profiles can bias prototype construction, a risk that hybrid fusion and abstention mitigate but do not eliminate. Unlike constrained or label-guided projection methods \cite{xia2023interactive, meng2024class}, our method externalizes analyst-defined seed groups as language-articulated semantic profiles and uses prototype-based embedding adjustments to steer projections without retraining. Future work should evaluate robustness under noisy or ambiguous seeds, compare against additional semantic-steering baselines \cite{bian2021deepsi, xia2023interactive}, extend to broader multimodal datasets, and study interactive inspection and refinement of prototypes with human analysts \cite{endert2012semantics}.

\section{Conclusion}
We presented a scalable semantic steering method that shifts LLM reasoning from individual items to user-defined groups. By generating group-level semantic profiles once and combining them with seed centroids into hybrid prototypes, the method propagates analyst intent through embedding-space operations without repeated LLM calls or retraining. The LitCovid results show that this group-level formulation can achieve global alignment comparable to per-item LLM reasoning while substantially reducing the cost of semantic propagation. The image case study further suggests that the same prototype-based mechanism can generalize beyond text. These findings indicate that group-level representations provide a practical path toward scalable LLM-augmented semantic steering.

\section*{Supplemental Materials}
\label{sec:supplemental_materials}
Supplemental materials are available on OSF at \url{https://osf.io/vr4e9/}. They include additional LitCovid projection comparisons, parameter sensitivity results, and the generated LitCovid group profiles.

% Refer to the instructions for this section (\cref{sec:supplement_inst}).
% Below is an example you can follow that includes the actual supplemental material for this template:

% All supplemental materials are available on OSF at \url{https://doi.org/10.17605/OSF.IO/2NBSG}, released under a CC BY 4.0 license.
% In particular, they include (1) Excel files containing the data for and analyses for creating \cref{tab:vis_papers} and \cref{fig:vis_papers}, (2) figure images in multiple formats, and (3) a full version of this paper with all appendices.
% Our other code is intellectual property of a corporation---Starbucks Research---and there is no feasible way to share it publicly.

% %% if specified like this the section will be committed in review mode
\acknowledgments{
This research was supported by industry, government, and institute members of the NSF SHREC Center, which was founded in the IUCRC program of the National Science Foundation. ChatGPT (OpenAI) was used for language editing and writing refinement of the manuscript. The authors take full responsibility for the content of the final manuscript.}

\bibliographystyle{abbrv-doi}

\bibliography{template}
\end{document}